\documentclass[twocolumn]{aastex61}

\received{2017 June 5}
\revised{2017 July 25}
\accepted{2017 August 2}
\submitjournal{ApJL}

\shorttitle{A millisecond pulsar discovery with LOFAR}
\shortauthors{Pleunis et al.}

\begin{document}

\title{A millisecond pulsar discovery in a survey of unidentified \textit{Fermi} $\gamma$-ray sources with LOFAR}

\correspondingauthor{Z. Pleunis}
\email{ziggy.pleunis@physics.mcgill.ca}

\author{Z. Pleunis}
\affiliation{Department of Physics and McGill Space Institute, McGill University, 3600 rue University, Montr\'{e}al, QC H3A 2T8, Canada}
\affiliation{Anton Pannekoek Institute for Astronomy, University of Amsterdam, Science Park 904, 1098 XH Amsterdam, The Netherlands}

\author{C.~G. Bassa}
\affiliation{ASTRON, the Netherlands Institute for Radio Astronomy, Postbus 2, 7990 AA Dwingeloo, The Netherlands}

\author{J.~W.~T. Hessels}
\affiliation{Anton Pannekoek Institute for Astronomy, University of Amsterdam, Science Park 904, 1098 XH Amsterdam, The Netherlands}
\affiliation{ASTRON, the Netherlands Institute for Radio Astronomy, Postbus 2, 7990 AA Dwingeloo, The Netherlands}

\author{V.~I. Kondratiev}
\affiliation{ASTRON, the Netherlands Institute for Radio Astronomy, Postbus 2, 7990 AA Dwingeloo, The Netherlands}
\affiliation{Astro Space Centre, Lebedev Physical Institute, Russian Academy of Sciences, Profsoyuznaya Str. 84/32, Moscow 117997, Russia}

\author{F.~Camilo}
\affiliation{SKA South Africa, Pinelands 7405, South Africa}

\author{I.~Cognard}
\affiliation{Laboratoire de Physique et Chimie de l'Environnement et de l'Espace, Universit\'e d'Orl\'eans/CNRS, F-45071 Orl\'eans Cedex 02, France}
\affiliation{Station de radioastronomie de Nan\c{c}ay, Observatoire de Paris, CNRS/INSU, F-18330 Nan\c{c}ay, France}

\author{J.-M.~Grie{\ss}meier}
\affiliation{Laboratoire de Physique et Chimie de l'Environnement et de l'Espace, Universit\'e d'Orl\'eans/CNRS, F-45071 Orl\'eans Cedex 02, France}
\affiliation{Station de radioastronomie de Nan\c{c}ay, Observatoire de Paris, CNRS/INSU, F-18330 Nan\c{c}ay, France}

\author{B.~W. Stappers}
\affiliation{Jodrell Bank Centre for Astrophysics, School of Physics and Astronomy, University of Manchester, Manchester M13 9PL, UK}

\author{A.~S. van Amesfoort}
\affiliation{ASTRON, the Netherlands Institute for Radio Astronomy, Postbus 2, 7990 AA Dwingeloo, The Netherlands}

\author{S. Sanidas}
\affiliation{Anton Pannekoek Institute for Astronomy, University of Amsterdam, Science Park 904, 1098 XH Amsterdam, The Netherlands}

\begin{abstract}

Using LOFAR, we have performed a very-low-frequency (115$-$155\,MHz) radio survey for millisecond pulsars (MSPs). The survey targeted 52 unidentified \textit{Fermi} $\gamma$-ray sources.  Employing a combination of coherent and incoherent dedispersion, we have mitigated the dispersive effects of the interstellar medium while maintaining sensitivity to fast-spinning pulsars. Toward 3FGL~J1553.1+5437 we have found PSR~J1552+5437, the first MSP to be discovered (through its pulsations) at a radio frequency $<$ 200\,MHz.  PSR~J1552+5437 is an isolated MSP with a 2.43 ms spin period and a dispersion measure of 22.9\,pc\,cm$^{-3}$. The pulsar has a very steep radio spectral index ($\alpha < -$2.8 $\pm$ 0.4). We obtain a phase-connected timing solution combining the 0.74\,years of radio observations with $\gamma$-ray photon arrival times covering 7.5\,years of \textit{Fermi} observations. We find that the radio and $\gamma$-ray pulse profiles of PSR~J1552+5437 appear to be nearly aligned. The very steep spectrum of PSR~J1552+5437, along with other recent discoveries, hints at a population of radio MSPs that have been missed in surveys using higher observing frequencies. Detecting such steep spectrum sources is important for mapping the population of MSPs down to the shortest spin periods, understanding their emission in comparison to slow pulsars, and quantifying the prospects for future surveys with low-frequency radio telescopes like SKA-Low and its precursors.

\end{abstract}


\section{Introduction} \label{sec:intro}

The Large Area Telescope \citep[LAT;][]{2009ApJ...697.1071A} on board the \textit{Fermi Gamma-ray Space Telescope} has (in)directly been responsible for dozens of millisecond pulsar (MSP) discoveries\footnote{See \url{http://tinyurl.com/fermipulsars} for an overview.} since it began operations in 2008 \citep{2013ApJS..208...17A}. Blind pulsation searches for MSPs in \textit{Fermi} data are possible \citep{2012Sci...338.1314P}, but have limited sensitivity due to the low count rate of $\gamma$-ray photons, and are furthermore computationally intensive and require a priori knowledge of orbital parameters to search for MSPs in binaries. Complementary targeted radio surveys of unidentified \textit{Fermi} sources have so far identified well over fifty radio-loud $\gamma$-ray MSPs by first detecting pulsed radio emission and later applying the timing model derived from radio observations to detect $\gamma$-ray pulsations \citep[e.g.][]{2012arXiv1205.3089R, 2015ApJ...810...85C}.  One-third of the 3033 $\gamma$-ray sources in the latest point-source catalog (3FGL) remain unidentified \citep[most are likely blazars, though certainly some of these are undiscovered MSPs;][]{2015ApJS..218...23A}, indicating a clear need for continued multi-wavelength follow-up observations.

Almost all MSP surveys to date have been performed at observing frequencies of 300$-$2000\,MHz and higher, thereby potentially missing MSPs with very steep spectra ($\alpha < -$3, where $S \propto \nu^\alpha$) and low flux densities.  

Here, we present the results of a Low-Frequency Array (LOFAR) pilot survey at 115$-$155\,MHz, targeted at \textit{Fermi} $\gamma$-ray sources. The primary goal of the survey was to test the hypothesis that very-steep-spectrum radio MSPs have been missed in previous pulsar surveys (both targeted and all-sky). This is important, e.g., for determining whether MSPs and slow pulsars have similar spectral index distributions -- a key observable related to the underlying emission mechanism \citep[e.g.][]{2013MNRAS.431.1352B}.  Such searches are also motivated by the hypothesis that the fastest-spinning MSPs are also preferentially steeper spectrum.  Mapping the MSP spin distribution is important for understanding the pulsar recycling mechanism, and probing beyond the currently known highest spin rate of 716\,Hz \citep{2006Sci...311.1901H} could reach the regime where rotation-derived neutron star radius limits become constraining.  We outline our survey strategy, observations and analysis in Section \ref{sec:lofar_survey}. In Section \ref{sec:j1552} we present the results. We discuss the results and conclude in Section \ref{sec:discussion}.

\section{LOFAR Survey of Unidentified $\gamma$-Ray Sources} \label{sec:lofar_survey}

\begin{figure}
\centering
\includegraphics[width=\columnwidth]{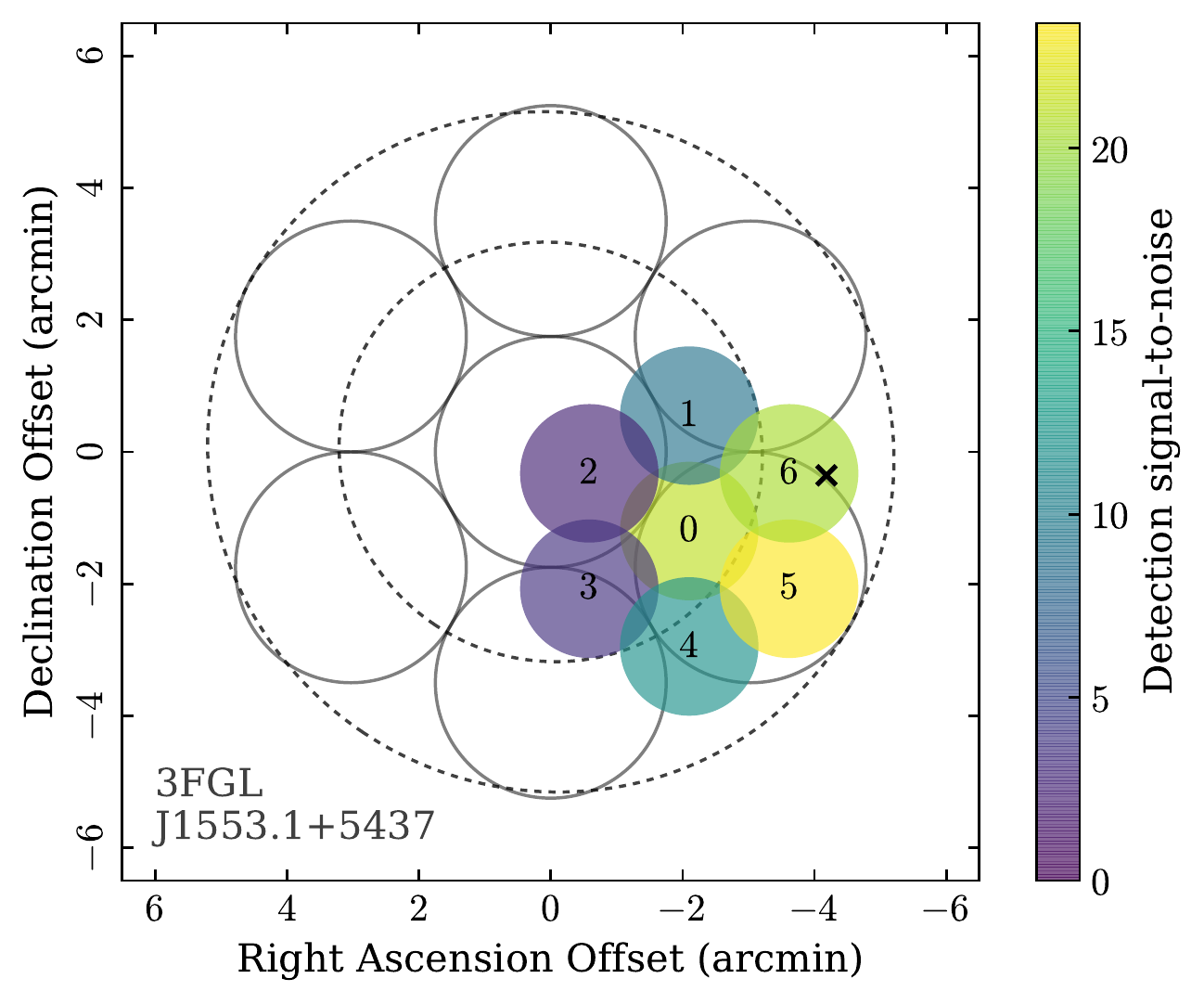}
\caption{Schematic representation of LOFAR tied-array beam positions for the observations of 3FGL sources. Gray open circles indicate the FWHM of the beams in the search observations. Real beams have side lobes and are elongated for non-zero zenith angles. The 68\% and 95\% confidence error ellipses from the third \textit{Fermi} point-source catalog are depicted with gray dashes. The filled circles represent the beams from the confirmation observation of PSR J1552+5437, with the color indicating the signal-to-noise of the folded pulsar signal. The confirmation observation used all LOFAR Core stations, and thus has a higher sensitivity in the center of the beams than the discovery observation. The pulsar's best-fit position from radio timing is denoted with a black cross.}
\label{fig:psr_pos}
\end{figure}

\begin{deluxetable*}{cccccc}
\centering
\tablecaption{LOFAR survey of unidentified \textit{Fermi}-LAT sources: source information. \label{tab:obs}}
\tablecolumns{6}
\tablewidth{0pt}
\tablehead{
\colhead{Name} & \colhead{Observation Epoch} & \colhead{Altitude\tablenotemark{a}} & \colhead{Azimuth\tablenotemark{a}} & \colhead{$r_{95}$\tablenotemark{b}} & Max. Gal. DM\tablenotemark{c} \\
\colhead{} & \colhead{(MJD)} & \colhead{($\degr$)} & \colhead{($\degr$)} & \colhead{(\arcmin)} & (pc cm$^{-3}$)
}
\startdata
3FGL J0017.1+1445 & 57376 & 51.9 & 184.5 & 5.08 & 37, 28\\ 
3FGL J0020.9+0323 & 57376 & 40.2 & 189.5 & 3.46 & 33, 23\\ 
3FGL J0031.6+0938 & 57376 & 45.1 & 201.3 & 5.67 & 35, 25\\ 
3FGL J0032.5+3912 & 57376 & 74.9 & 209.5 & 5.3 & 61, 55\\ 
3FGL J0102.1+0943 & 57376 & 45.7 & 198.1 & 4.93 & 35, 25\\ 
\enddata
\tablenotetext{a}{At the midpoint of the LOFAR observation.}
\tablenotetext{b}{Semimajor axis of 95\% confidence error region in the 3FGL catalog.}
\tablenotetext{c}{According to the NE2001 \citep{2002astro.ph..7156C} and the YMW16 \citep{2017ApJ...835...29Y} models for Galactic electron density, respectively.}
\tablecomments{Table \ref{tab:obs} is published in its entirety in the machine-readable format. A portion is shown here for guidance regarding its form and content.}
\end{deluxetable*}

\subsection{Survey Setup}

We have used the LOFAR High Band Antennas \citep[HBAs;][]{2013A&A...556A...2V} of 21 of the 24 LOFAR Core stations\footnote{We excluded station CS013 because it had a $45\degr$ dipole rotation error at the time of the observations and the two outermost stations CS103 and CS302 to be able to cover the error ellipses of a larger number of \textit{Fermi} sources.} to form 7 tied-array beams \citep{2011A&A...530A..80S}. This observational setup has baselines up to 2.3\,km and provides tied-array beams of $\sim$~3\farcm5 in diameter (FWHM) at the central frequency, with 7 beams covering a total circular field-of-view (FoV) of about 10\arcmin~in diameter (see Fig.~\ref{fig:psr_pos}). With this setup we have observed 52 out of 1010 unidentified $\gamma$-ray sources from the 3FGL \textit{Fermi}-LAT point-source catalog \citep{2015ApJS..218...23A}. These 52 sources were selected as they are visible to LOFAR (source elevation $>$ 30\degr~during transit), located away from the Galactic plane ($|b| > 10\degr$; where the sky temperature and scattering at 135 MHz are significantly lower), and because they have positional uncertainty regions less than 10\arcmin~in diameter (i.e. fit the FoV of 7 tied-array beams). No cuts on the spectral parameters of the sources were performed. The observed sources and some of their parameters are listed in Table~\ref{tab:obs}. The sample of \textit{Fermi} sources searched here does not overlap with that of \citet{2017ApJLsubmitted}.

We employed a semi-coherent dedispersion scheme, aimed at mitigating the effects of dispersive smearing and implemented in \texttt{cdmt} \citep{2017A&C....18...40B}. To allow coherent dedispersion, we have recorded complex voltage data for dual-polarization, Nyquist sampled subbands of 195.3125\,kHz bandwidth (5.12\,$\mu$s sampling). To maximize sensitivity and FoV we have used signals from 200 subbands in the 115$-$155\,MHz frequency range (39.06\,MHz bandwidth). Modest integration times of $T_\mathrm{obs} =$ 20 minutes were chosen to maintain sensitivity to accelerated signals from binary systems. 

For each observation, the 200 frequency subbands were coherently dedispersed to 80 evenly spaced trial dispersion measures (DMs), ranging from 0.5 to 79.5\,pc\,cm$^{-3}$ (about twice the expected maximum Galactic DM for most of the surveyed sources), and channelized into a total of 1600 channels, using \texttt{cdmt}. The time and spectral resolution after channelization were 40.96\,$\mu$s and 24.41\,kHz, respectively. Around each coherent DM trial we made incoherent DM trials in steps of 0.002\,pc\,cm$^{-3}$. The two DM step sizes are chosen to limit the total (intra-channel and $\Delta$DM) dispersive smearing compared to the true DM of the source to a maximum of 0.15\,ms (see the top panel in Fig.~\ref{fig:sensitivity}). Each dedispersed time series was searched for accelerated periodic signals in the frequency domain, and the 200 best pulsar candidates from each beam, according to a modified version of PRESTO's \texttt{accel\_sift.py} sifting script \citep{2001PhDT.......123R}, were folded and inspected by eye. 

Confirmation observations used all LOFAR Core stations (baselines up to 3.5 km), and thus have tied-array beams with a $\sim$~3 times smaller area of $\sim$~2\arcmin~in diameter (FWHM) at the central frequency. Furthermore, the tied-array ring size (the offset of the center of the outer beams from the center of the pointing) is reduced to 1\farcm75, such that the beams overlap slightly, and the position of a newly discovered pulsar can be refined by weighting the signal-to-noise ratios of detections in the different beams (see the colored filled circles in Fig.~\ref{fig:psr_pos} for an illustration). However, note that the ionosphere can shift beams by approximately an arcminute during periods of strong ionospheric turbulence.  This can somewhat reduce the accuracy of this positional determination method.

\subsection{Survey Sensitivity}

\begin{figure}
\centering
\includegraphics[width=\columnwidth]{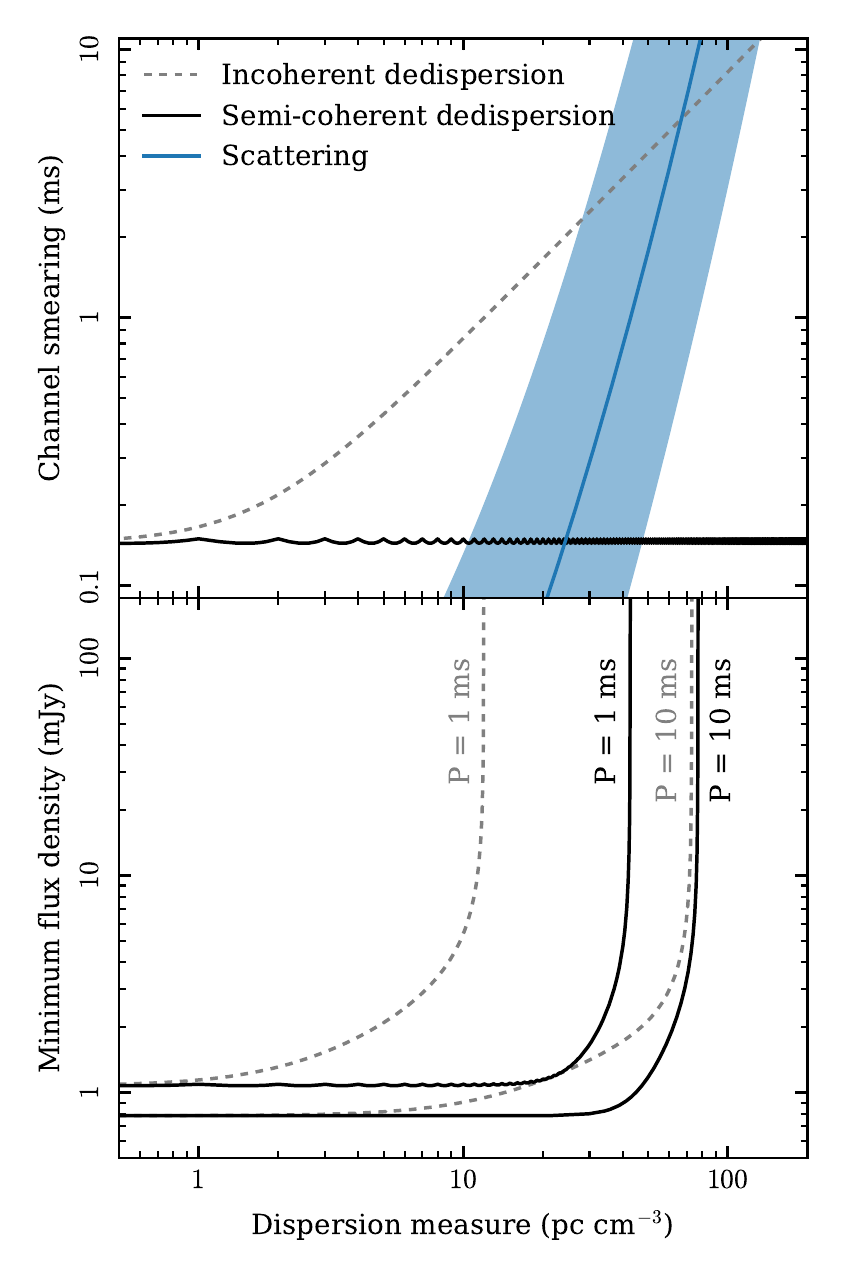}
\caption{LOFAR targeted survey sensitivity to MSPs at a central observing frequency of 135 MHz. \textit{Top}: the leftover dispersive channel smearing is depicted for incoherent (gray dashed line) and semi-coherent (black solid line; with the setup described in \S \ref{sec:lofar_survey}) dedispersion. The expected scattering time based on \citet{2004ApJ...605..759B} is depicted with a blue solid line, with the blue shaded region showing up to 10$\times$ smaller and larger values, to reflect the scatter in the relation. For DMs $\gtrsim$ 25 pc cm$^{-3}$ scattering becomes the dominant source of smearing within channels and scattering starts to drastically reduce the sensitivity to MSPs for DMs $\gtrsim$ 50 pc cm$^{-3}$. \textit{Bottom}: the minimum flux density an MSP needs to have at 135 MHz in order to be discovered by an incoherent (gray dashed lines) and by a semi-coherent (black solid lines) search pipeline with the effect of scattering taken into account.}
\label{fig:sensitivity}
\end{figure}

Although the effects of dispersive smearing within a channel can be mitigated by the use of coherent dedispersion, the sensitivity of any pulsar survey at low radio frequencies is ultimately limited by scattering (approximately $\propto \nu_\mathrm{obs}^{-4}$), which results in an exponential broadening of the observed pulse shapes. We calculated the expected scattering times using the empirical fit for scattering as a function of DM made by \citet{2004ApJ...605..759B}, and compared this to dispersive smearing within channels (see the top panel in Fig.~\ref{fig:sensitivity}). We have calculated the minimum detectable flux density our survey was sensitive to using the modified radiometer equation for pulsars \citep[][Appendix 1.4]{2012hpa..book.....L}, where we have used $\sigma$ = 10 as the minimum signal-to-noise ratio for a convincing pulsar candidate (although candidates with a somewhat lower signal-to-noise ratio were also investigated), $\beta \approx$ 1.0 as the digitization correction factor (survey observations were processed with 8-bit integer bit depth), $T_\mathrm{sys} \approx$ 400 K as the temperature of the telescope and the sky at the observing frequency, and $G \approx$ 5.6 K Jy$^{-1}$ as the telescope's gain\footnote{The gain $G \sim T_\mathrm{sys}/$SEFD$_\mathrm{Core} \approx 400$ K/(3000 Jy/42) $\approx$ 5.6 K Jy$^{-1}$ when using 21 of the LOFAR Core stations \citep{2013A&A...556A...2V}; here, SEFD is the system equivalent flux density.}. Sensitivity curves for a pulsar with a 1 and a 10 ms spin period and an intrinsic 10\% duty cycle are shown in the bottom panel of Figure \ref{fig:sensitivity}. Out to DMs of about 40 pc cm$^{-3}$ we were sensitive to 2 ms pulsars brighter than $\sim$~2 mJy, if the source was not eclipsed at the time of observation (many binary $\gamma$-ray MSPs are eclipsed for up to $\sim$~50\% of their orbit). This flux limit applies to observations at zenith; the sensitivity falls off approximately as $\sin^{-1.4}(\theta_\mathrm{z})$, where $\theta_\mathrm{z}$ is the zenith angle \citep{2015A&A...576A..62N}.

\section{Discovery and Timing of PSR J1552+5437}\label{sec:j1552}

\begin{figure}
\centering
\includegraphics[width=\columnwidth]{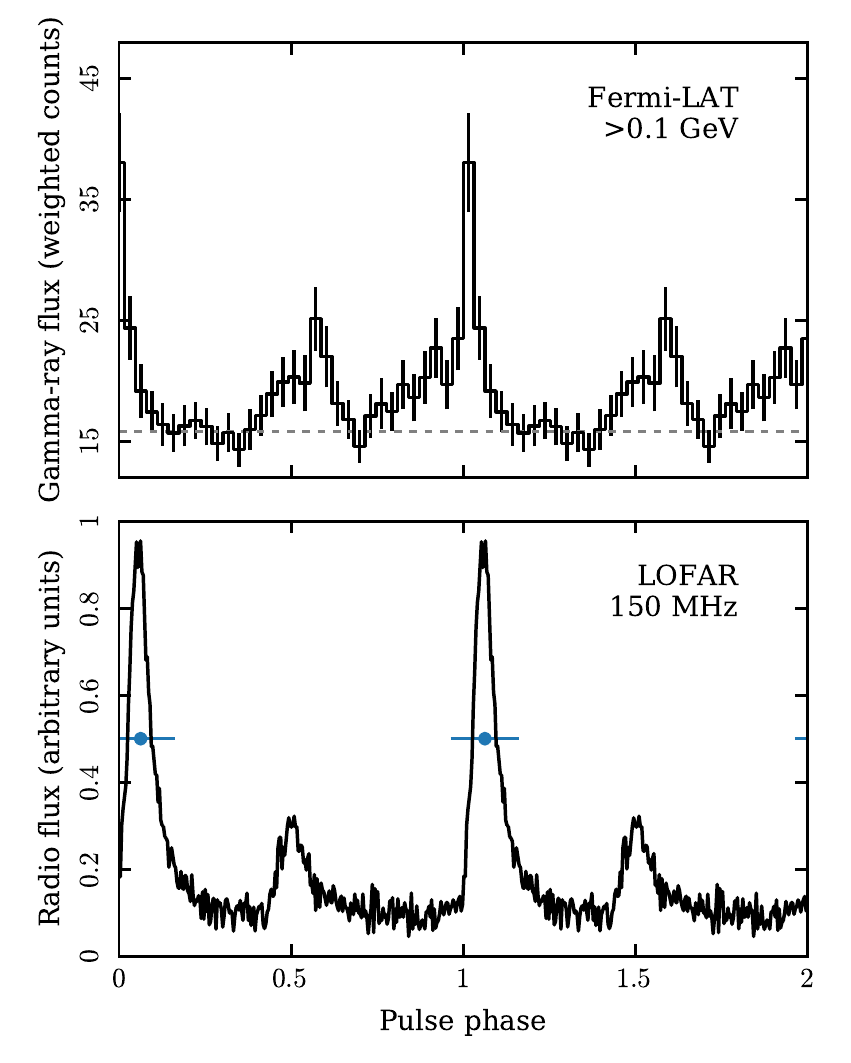}
\caption{Aligned $\gamma$-ray and radio pulse profiles of PSR J1552+5437. The $\gamma$-profile contains $\sim$~7.5 years of \textit{Fermi} photons, weighted with their probability of coming from the source, and folded in 32 phase bins. The errors on the phase bins as well as the background (gray dashed line) are estimated as in \citet{2013ApJS..208...17A}. The radio profile is a stacked pulse profile for 19 timing observations (total integration time 6\,hours) at a central frequency of 150 MHz, folded in 256 bins. The blue error bars indicate the potential radio profile phase shift due to a DM variation of 10$^{-3}$ pc cm$^{-3}$ over the course of the \textit{Fermi} mission.}
\label{fig:psr_prof}
\end{figure}

We discovered an isolated pulsar with a 2.43 ms spin period at a DM of 22.9 pc cm$^{-3}$, in a pointing toward 3FGL J1553.1+5437 (see Fig.~\ref{fig:psr_prof} for its radio and $\gamma$-ray pulse profile). The pulsar was detected in two adjacent beams in the discovery observation.

\subsection{Radio Analysis}

\begin{figure*}
\centering
\includegraphics[width=\textwidth]{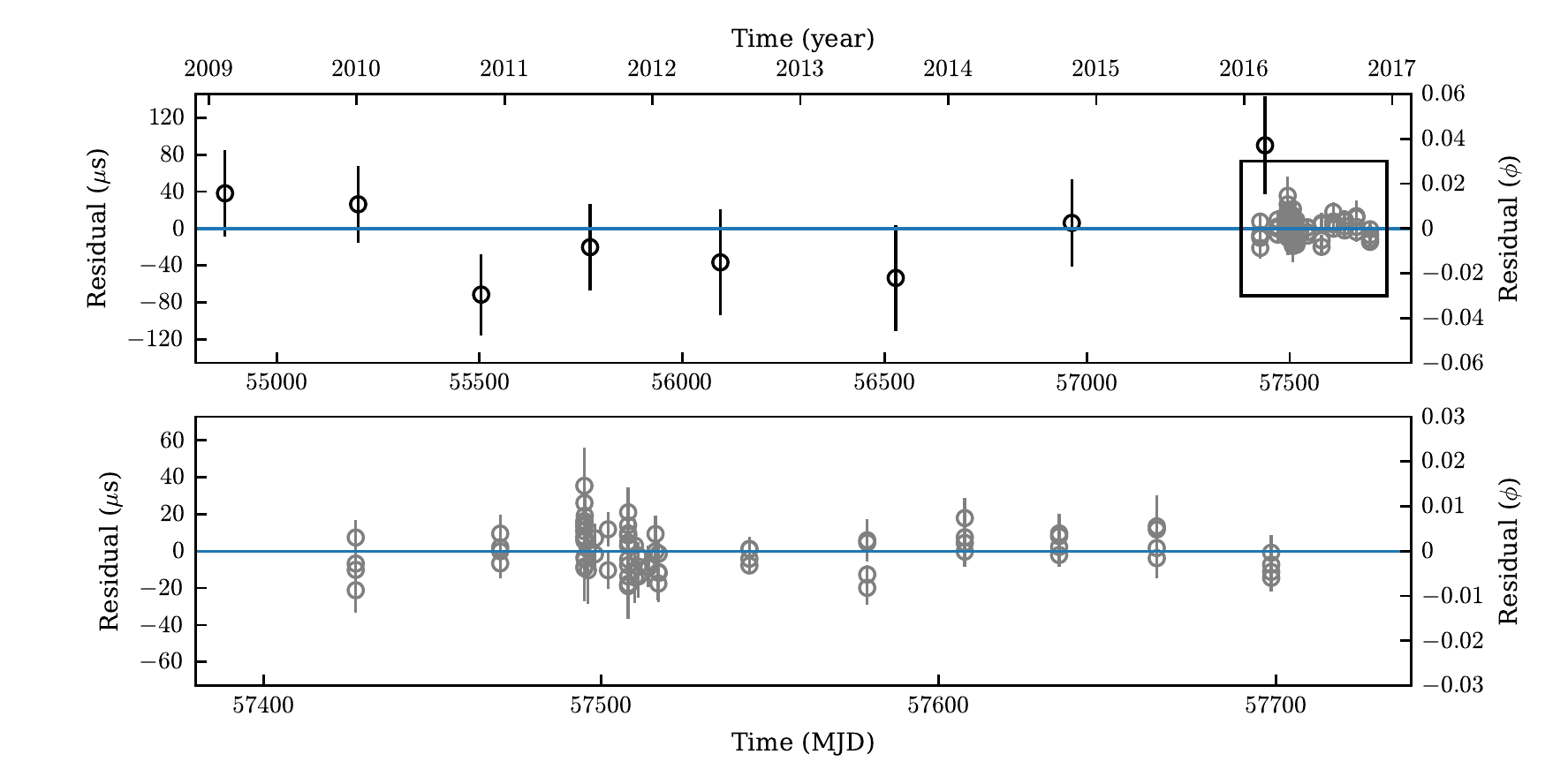}
\caption{Timing residuals for PSR J1552+5437 as a function of time. The model is depicted with a blue line, the \textit{Fermi} TOAs with black circles, and the LOFAR TOAs with gray circles. The lower panel is a magnification of the upper panel, showing only the LOFAR TOAs.}
\label{fig:psr_toas}
\end{figure*}

Following the discovery and confirmation of the pulsar we started a timing campaign with LOFAR. Timing observations use all Core stations and the HBA bandwidth from 110 to 188\,MHz, two times wider than possible in the survey observations. After initial dense and logarithmically spaced 10 minute observations spanning two weeks, the pulsar was observed once per month for 20 minutes. All observations are dedispersed and folded using the LOFAR Pulsar Pipeline \citep[e.g.][]{2016A&A...585A.128K}. Pulse times-of-arrival (TOAs) are extracted from 5 minute sub-integrations using tools from the PSRCHIVE\footnote{\url{http://psrchive.sourceforge.net}} \citep{2004PASA...21..302H} pulsar software package. We have used TEMPO2\footnote{\url{http://sourceforge.net/projects/tempo2/}} \citep{2006MNRAS.369..655H} to obtain an initial phase-connected timing solution spanning 0.74 years and fitted for position, spin frequency, and DM (see Fig.~\ref{fig:psr_toas}). The \texttt{efac/equad} plug-in \citep{2015MNRAS.446.1657W} was used to rescale the LOFAR TOA uncertainties, suggesting that a multiplication factor of 1.3 and an additional uncertainty of 0.8 $\mu$s (multiplied with and added to the initial uncertainty in quadrature) better reflect the expected Gaussian scatter of the residuals. Note that scattering can influence the measured DM and that there are thus likely systematic uncertainties on the DM that are larger than the nominal TEMPO2 error listed in Table \ref{tab:timing}. Also, the frequency dependence of the pulse profile might bias the measured DM value.

\begin{deluxetable}{ll}
\centering
\tablecaption{Parameters for PSR J1552+5437}
\tablehead{\colhead{Parameter} & \colhead{Value}}
\startdata
\multicolumn{2}{c}{Timing Parameters (Radio and $\gamma$-Ray)} \\
\hline
Right ascension (J2000)  & 15$^\mathrm{h}$52$^\mathrm{m}$53$\fs$33117(17) \\ 
Declination (J2000)  & +54$\degr$37\arcmin05$\farcs$7866(14) \\ 
Spin frequency (Hz)  & 411.88053142429(10) \\ 
Frequency derivative (Hz s$^{-1}$)  & $-$4.746(17) $\times$ 10$^{-16}$ \\
Dispersion measure (pc cm$^{-3}$)  & 22.9000(5) \\ 
Span of timing data (MJD)  & 54871.7--57698.5 \\
Epoch of timing solution (MJD) & 56285 \\ 
Number of TOAs & 88 \\
RMS timing residual ($\mu$s) & 10.1 \\
Reduced $\chi^2$ value & 1.1 \\
Clock correction procedure & TT(BIPM2011) \\
Solar system ephemeris model & DE421 \\
\hline
\multicolumn{2}{c}{Radio Flux Densities} \\ 
\hline 
Flux density at 150 MHz (mJy) & 3.8 $\pm$ 1.9 \\
Flux density at 820 MHz ($\mu$Jy) & $<$ 17 \\
Flux density at 1.4 GHz ($\mu$Jy) & $< $ 20 \\
\hline
\multicolumn{2}{c}{Derived Parameters} \\
\hline 
Spin period (ms) & 2.4279 \\
Spectral index & $< -$2.8 $\pm$ 0.4 \\
Galactic longitude ($\degr$) & 85.6 \\
Galactic latitude ($\degr$) & 47.2 \\
DM-derived distance\tablenotemark{a} (kpc) & 1.2, 2.6 \\
Spin-down luminosity\tablenotemark{a,b} (erg s$^{-1}$) & (8.9, 9.4) $\times$ 10$^{33}$ \\
Surface magnetic field\tablenotemark{a,b} (G) & (9.0, 9.2) $\times$ 10$^7$ \\
Characteristic age\tablenotemark{a,b} (years) & (1.2, 1.1) $\times$ 10$^{10}$ \\
\hline
\multicolumn{2}{c}{$\gamma$-Ray Parameters} \\
\hline 
$\gamma$-ray-radio profile lag ($\phi$) & 0.042 $\pm$ 0.004 $\pm$ 0.1 \\
$\gamma$-ray peak separation ($\phi$) & 0.53 $\pm$ 0.01 \\
$\gamma$-ray photon index  & 1.4 $\pm$ 0.3 \\
$\gamma$-ray cutoff energy (GeV)  & 3.7 $\pm$ 1.6 \\
Photon flux (cm$^{-2}$ s$^{-1}$)  & (2.5 $\pm$ 0.9) $\times$ 10$^{-9}$ \\
Energy flux (erg cm$^{-2}$ s$^{-1}$)  & (2.7 $\pm$ 0.4) $\times$ 10$^{-12}$ \\ 
Luminosity\tablenotemark{a} (10$^{32}$ erg s$^{-1}$) & (4.7 $\pm$ 0.7), (22 $\pm$ 3.2) \\
Efficiency\tablenotemark{a} (\%) & 6.1, 28.6 \\
\enddata
\tablenotetext{a}{Based on the NE2001 \citep{2002astro.ph..7156C} and the YMW16 \citep{2017ApJ...835...29Y} models, respectively.}
\tablenotetext{b}{Upper limit: corrected for the acceleration due to the kinematics of the Galaxy, but not for the Shklovskii effect.}
\label{tab:timing}
\end{deluxetable}

The pulsar's flux density was measured in all timing observations by calibrating the observations using an improved Hamaker beam model \citep{2006A&A...456..395H}, and comparing the on-pulse with the off-pulse window \citep[full details of LOFAR MSP flux calibration are described by][]{2016A&A...585A.128K}. These measurements lead to a mean flux density for 19 observations at 150 MHz of 3.8 $\pm$ 1.9 mJy (50\% uncertainty), but the observed flux density can vary by a factor $\sim$~2 from observation to observation -- possibly because of refractive scintillation, though RFI and ionospheric beam jitter can also influence this. A search for the Faraday rotation measure toward the pulsar using PSRCHIVE's \texttt{rmfit} routine did not converge for any of the LOFAR observations, likely because the pulsar shows little or no polarization beyond the detection limit. 

PSR J1552+5437 was observed at L-band for $9\times30$\,min and $4\times1$\,hour with the Lovell (400\,MHz bandwidth at 1532\,MHz center frequency) and Nan\c{c}ay (NRT; 512\,MHz at 1486\,MHz) radio telescopes, but not detected, limiting the flux density to less than 13\,$\mu$Jy, when these data sets are co-added (under the assumption that diffractive scintillation is averaged out; here, the expected scintillation bandwidth is only $\sim$~10 MHz at 1400 MHz). A non-detection of the pulsar at 820 MHz (200\,MHz bandwidth) in a 1.5-hour observation using the Robert C. Byrd Green Bank Telescope sets an upper limit to the pulsar's flux density at 820 MHz of 17 $\mu$Jy. These upper limits are calculated using the modified radiometer equation for pulsars \citep[][Appendix 1.4]{2012hpa..book.....L}, assuming that PSR J1552+5437 would have been detected in those bands if it had a signal-to-noise ratio of at least 5 and a similar pulse width to our LOFAR detections. Based on the same assumptions, the upper limit for NRT is confirmed to be 20\,$\mu$Jy after flux calibration of the $4\times1$\,hour of observation with a pulsed noise diode and a calibration source. The detections at 150 MHz and the upper limits at 820 and 1400\,MHz constrain the radio power-law spectral index of PSR J1552+5437 to be $\alpha < -$3.2 $\pm$ 0.4, where we assumed that the radiometer equation has a 50\% uncertainty. A more conservative upper limit, however, takes into account the potential overestimate of LOFAR pulsar fluxes by a factor $\sim$~2 that was noted by \citet{2016ApJ...829..119F}. In that case the LOFAR 150\,MHz flux density would be 1.9 $\pm$ 0.9 mJy, and the upper limit on the spectral index $\alpha < -$2.8 $\pm$ 0.4. Future observations at 350\,MHz will also be useful for mapping the spectrum.

We have also observed the pulsar with LOFAR's Low Band Antennas (LBAs) for 1 hour at $30-90$\,MHz on MJD 57496, but were unable to detect the pulsar by folding the data with the best-fit parameters derived from the timing analysis. This is unsurprising given the faintness of the source in the LOFAR HBA and the increased system temperature $T_\mathrm{sys}$ in the LBA. Also, the scattering tail that is already visible in the radio profile at 150 MHz (Fig.~\ref{fig:psr_prof}) will be $\sim 10 \times$ larger in the LBA range and would smear out the pulsations. Only three (very bright, and unscattered) MSPs have so far been detected using the LOFAR LBAs \citep{2016A&A...585A.128K}.

\subsection{$\gamma$-Ray Analysis}

We downloaded the \textit{Fermi}-LAT Pass 8 photons of the \texttt{SOURCE} class from 2008 August 4 (the start of the mission) to 2016 October 14, within 20$\degr$ of the best position derived from radio timing. Using the \textit{Fermi} Science Tools, we selected the photons in the energy range 0.1--100 GeV using the recommended cuts. We performed a binned maximum likelihood \texttt{gtlike} analysis on the photons in the 20\degr~$\times$~20\degr~square centered on the timing position, leaving only the spectral parameters of the sources within the inner 5\degr~radius free. Our source model was based on the 3FGL catalog and as models for the Galactic diffuse emission and isotropic diffuse background we used the \texttt{gll\_iem\_v06.fits} \citep{2016ApJS..223...26A} and \texttt{iso\_P8R2\_SOURCE\_V6\_v06.txt} templates\footnote{\url{https://fermi.gsfc.nasa.gov/ssc/data/access/lat/BackgroundModels.html}} respectively. 3FGL J1553.1+5437 moved to the pulsar's timing position is detected with a test statistic TS value of 205 (about 14$\sigma$, while the source had a $\sim$~8.5$\sigma$ significance previously) using an exponentially cutoff power-law model to describe its spectrum. The exponentially cutoff model is preferred over a simpler power law as TS$_\mathrm{cut} \equiv 2 \Delta$log(likelihood) = 15 $>$ 9 \citep[likelihood ratio test, following][]{2013ApJS..208...17A}, and the best-fit parameters are listed in Table \ref{tab:timing}.

Based on the spectral analysis, all the events in the region around the source were assigned a probability of originating from 3FGL J1553.1+5437 using \texttt{gtsrcprob} \citep{2011ApJ...732...38K}. Selecting only those events with a probability $> 20$\% resulted in 350 photons. Pulsar rotational phases $\phi_i(t)$ were computed based on the radio timing solution using TEMPO2\footnote{\url{http://sourceforge.net/projects/tempo2/}} \citep{2006MNRAS.369..655H} with the \texttt{fermi} plug-in \citep{2011ApJS..194...17R}. Folding the $\gamma$-ray photons over the range where the radio timing solution was valid did not result in a significant pulse profile, and we thus performed a brute-force search over the pulsar's spin frequency $f$ and spin-frequency derivative $\dot{f}$ to find a coherent solution over the 7.5 years of \textit{Fermi} data (neglecting higher order effects in this search is feasible because the MSP is likely isolated).

In the brute-force search, the barycentered phases were updated using the Taylor series
\begin{equation}
\phi_i(t) = \phi_{i,0} + f (t_i - t_0) + \frac{1}{2} \dot{f} (t_i - t_0)^2, 
\end{equation}
for 100 $\times$ 100 values of $f$ and $\dot{f}$ within two times the error range of the radio timing solution. The H-test \citep{1989A&A...221..180D} of the folded pulse profile was calculated for each trial. With the $f$ and $\dot{f}$ that maximized H to 70, it was possible to significantly fold all 350 \textit{Fermi} photons, which confirms the link between PSR J1552+5437 and 3FGL J1553.1+5437. 

To lift the degeneracy between astrometric and rotational parameters in the timing solution we included the $\gamma$-ray data in our timing analysis. We used an unbinned maximum likelihood method to extract 8 topocentered TOAs with at least a 3$\sigma$ detection from the 350 \textit{Fermi} photons \citep{2011ApJS..194...17R}. More sophisticated and sensitive unbinned methods for extracting $\gamma$-ray TOAs have been developed in recent years \citep[e.g.][]{2015ApJ...814..128K, 2015ApJ...807...18P}, but using the method described above suffices for our present purposes. 

The results of joint radio and $\gamma$-ray timing are listed in Table \ref{tab:timing}, the $\gamma$-ray profile folded with the final timing solution is depicted in Figure \ref{fig:psr_prof}, and the timing residuals as a function of time are shown in Figure \ref{fig:psr_toas}. The timing position is not at the center of the three beams with the best detections in the confirmation observations (see Fig.~\ref{fig:psr_pos}), while the timing position based on the radio data alone agrees with the full timing solution to within a few hundredths of an arcsecond. Possibly, the ionosphere has caused the beams to shift by $\sim$~1\arcmin~in the confirmation observation. We also note that the schematic shown in Figure \ref{fig:psr_pos} is only a rough approximation of the true beam shapes.

The observed spin period derivative of 2.80 $\times$ 10$^{-21}$ s s$^{-1}$ is not the intrinsic value, as it has to be corrected for the non-zero proper motion of the pulsar, the Shklovskii effect \citep{1970SvA....13..562S} and for movement due to the kinematics of the Galaxy \citep[e.g.][]{1995ApJ...441..429N}. The Galactic contribution is ($-$4.36, $-$6.20) $\times~10^{-22}$ s s$^{-1}$ for a (1.2, 2.6) kpc distance in the line-of-sight. This is the sum of the differential Galactic rotation and the $k_\mathrm{z}$ term. Adding this correction leads to a spin-frequency derivative of (3.24, 3.42) $\times$ 10$^{-21}$ s s$^{-1}$. With the current data, it was not possible to significantly fit for the proper motion of the pulsar. However, the uncertainty on the fit values limits the proper motion to $<$ 36.8 mas yr$^{-1}$ (3$\sigma$), corresponding to a Shklovskii correction to the spin period derivative of $<$ (9.6, 20.8) $\times~10^{-21}$ s s$^{-1}$. The inferred surface magnetic field strength based on the observed spin period and spin period derivative, corrected for Galactic acceleration, is with (9.0, 9.2) $\times$ 10$^7$ G already one of the lowest pulsar magnetic fields measured to date, and will become slightly lower after correcting $\dot{P}$ with an extended timing baseline.

Finally, we consider the offsets between the radio and $\gamma$-ray pulse peaks. As can be seen in Figure \ref{fig:psr_prof}, both the radio and the $\gamma$-ray profile show a main pulse and a subpulse offset by about half a rotational phase. The $\gamma$-ray profile does not show any additional features when the number of phase bins is increased. We have set the rotational phase $\phi =$ 0 at the onset of the main pulse of the LOFAR radio profile. To quantify the peak separations, we fitted a Gaussian profile to both radio pulses. The peak around phase 0 in the probability-weighted $\gamma$-ray profile was fitted using two Lorentzian profiles, and the other pulse with one Lorentzian profile, on top of the background. The maximum of the radio profile is at $\phi_\mathrm{r} =$ 0.063 $\pm$ 0.002 (where the rotational phase is defined between 0.0 and 1.0, and errors are statistical), the radio subpulse peaks at $\phi = $ 0.51 $\pm$ 0.01, and the peaks of the $\gamma$-ray profiles are at $\phi_1 =$ 0.021 $\pm$ 0.004 and $\phi_2 =$ 0.553 $\pm$ 0.013. Adopting this $\gamma$-ray peak definition leads to a radio-to-$\gamma$-ray lag of $\delta = \phi_1 - \phi_\mathrm{r} \simeq$ 0.04, and a $\gamma$-ray peak separation of $\Delta = \phi_2 - \phi_1 \simeq$ 0.53. These numbers seem consistent with other LAT MSPs \citep{2013ApJS..208...17A}; the $\Delta \gtrsim$ 0.5 in phase, however, might indicate that the definition of the first and second $\gamma$-ray peaks could, in principle, be reversed. If that is the case, the $\gamma$-rays either lead the radio by $\sim$~0.49 or trail it by $\sim$~0.51 in phase.

The alignment of the main peaks of the radio and $\gamma$-ray profile might be real, but it could also reflect the limited baseline of the radio timing of the pulsar. A 10$^{-3}$ pc cm$^{-3}$ variation in DM over the length of the \textit{Fermi} mission could lead to a shift of $\sim$~0.1 in rotational phase between the radio and the $\gamma$-ray profiles. Such a DM variation would be consistent with those seen for other MSPs \citep{2013MNRAS.429.2161K}. A higher-frequency radio profile (e.g. measured at 820 MHz) would be less sensitive to DM variations, but we have so far been unable to detect PSR J1552+5437 at higher radio frequencies.

\section{Discussion and outlook}\label{sec:discussion}

In our targeted LOFAR survey toward 52 unidentified \textit{Fermi}-LAT $\gamma$-ray sources, we discovered one MSP. The newly discovered MSP, PSR J1552+5437, has a low inferred magnetic field (B $<$ 9.2 $\times$ 10$^7$ G), and a very steep power-law radio spectrum ($\alpha < -$2.8 $\pm$ 0.4). Only 9 pulsars in the ATNF pulsar catalog\footnote{\url{http://www.atnf.csiro.au/research/pulsar/psrcat}} have lower inferred magnetic fields \citep{2005AJ....129.1993M}, and only 8 of the 200 GMRT-detected pulsars have spectral indices $< -$2.8 \citep{2016ApJ...829..119F}. In 2.9 ks of \textit{Swift}-XRT observations of 3FGL J1553.1+5437 no source is detected above 3$\sigma$\footnote{\url{http://www.swift.psu.edu/unassociated}} \citep{2013ApJS..207...28S}, making the pulsar a suboptimal target for, e.g., the \textit{NICER} mission \citep{2014SPIE.9144E..20A}, despite its relatively small DM-distance.

This pilot survey has shown that LOFAR is capable of discovering MSPs. In fact, it is the first digital aperture array to discover an MSP directly through its pulsed signal, and this is the lowest radio frequency (135\,MHz) at which any MSP has been discovered to date. As a follow-up survey of \textit{Fermi} unidentified sources, however, it has a success rate of only a few percent, which is low compared to similar surveys at higher frequencies \citep[which have success rates of 12 to 26\%; see][for an overview]{2016ApJ...819...34C}. A refined selection of \textit{Fermi} targets (choosing the most pulsar-like unidentified $\gamma$-ray sources) will likely increase the success rate of future LOFAR MSP searches, as suggested by the recent discovery of PSR J0952$-$0607 \citep{2017ApJLsubmitted}. Furthermore, instead of only once, each source should be observed two or three times to reduce the probability of catching the pulsar during an eclipse.

Nonetheless, 3FGL J1553.1+5437 is a relatively weak \textit{Fermi} point source with a relatively large positional uncertainty that was classified as a likely active galactic nucleus using machine-learning techniques \citep{2016ApJ...820....8S}. Targeted radio surveys often favor the bright and well-constrained $\gamma$-ray sources with weaker MSPs in unidentified \textit{Fermi} sources going unnoticed. This was also observed in a recent blind search for $\gamma$-ray pulsars in \textit{Fermi} data, where at least two pulsars were discovered close to or slightly outside the edge of the search region \citep{2017ApJ...834..106C}. This, and the fact that PSR J1552+5437 was not detected at 820\,MHz and 1.4\,GHz, advocates repeat searches of \textit{Fermi} unidentified sources -- even those that \textit{a priori} appear less pulsar-like -- using low-frequency radio telescopes and covering a reasonable region around the quoted positional uncertainty.

PSR~J1552+5437 shows characteristics also seen in other MSPs with nearly aligned radio and $\gamma$-ray profiles. For this class of MSPs, models have been developed where both the radio and $\gamma$-ray emission are produced close to the light cylinder, with the radio emission showing some linear polarization \citep{2004ApJ...606.1125D, 2012ApJ...744...34V}. In a study of 30 $\gamma$-ray MSPs, \citet{2013MNRAS.430..571E} indeed find that MSPs with aligned profiles have the largest inferred magnetic fields at their light cylinders\footnote{$B_\mathrm{lc} \propto P^{-5/2}\dot{P}^{1/2}$; about 5 $\times$ 10$^4$ G for PSR J1552+5437 and typically $>$ 10$^4$ for $\gamma$-ray MSPs, with PSR B1937+21 having $\sim$~10$^6$ G.}. They furthermore find that those MSPs have the steepest radio spectra, with a probability of less than 1\% of originating from the same spectral distribution as other ($\gamma$-ray) MSPs.

In modeling the light curves of $\gamma$-ray MSPs with a variety of magnetospheric models, \citet{2014ApJS..213....6J} find that the MSPs with radio and $\gamma$-ray peak alignment within 0.1 in rotational phase, are best fitted by outer gap and slot gap models \citep[see also][and references therein]{2015CRPhy..16..641G}. PSR J1552+5437 supports the hypothesis that to find the fastest-spinning pulsars we need to find the MSPs with the steepest spectra; further surveys with LOFAR and other low-frequency radio telescopes (as well as SKA-Low in the future) are instrumental in this quest.

In further support of this hypothesis, \citet{2016ApJ...829..119F} find that of the 16 pulsars with the steepest spectra (spectral index $< -$2.5) in spectral measurements of 200 GMRT-detected pulsars (at 150\,MHz; in the same frequency range as our LOFAR survey), 12 are MSPs, and all but one are $\gamma$-ray MSPs. A new detailed population study, similar to the ones performed by \citet{1998ApJ...501..270K} or \citet{2013MNRAS.431.1352B}, but including the findings of low-frequency surveys for MSPs, can establish whether the faster-spinning pulsars truly have steeper spectra on average, and whether the spectral distributions of slow pulsars and ($\gamma$-ray) MSPs diverge.

\acknowledgments

We thank Rob Archibald, Maura Pilia, David Smith, Joris Verbiest, the \textit{Fermi} LAT team, and the referee for useful discussions; LOFAR Science Operations and Support for their help in scheduling and effectuating these observations; and Andrew Lyne for obtaining observations with the Lovell Telescope. The research leading to these results has received funding from the European Research Council under the European Union's Seventh Framework Programme (FP7/2007-2013) / ERC grant agreement No. 337062 (DRAGNET; PI: Hessels). J.W.T.H. is an NWO Vidi fellow. LOFAR, the Low-Frequency Array designed and constructed by ASTRON, has facilities in several countries, that are owned by various parties (each with their own funding sources), and that are collectively operated by the International LOFAR Telescope (ILT) foundation under a joint scientific policy. Survey observations were taken under proposal LC5\_002 (PI: Pleunis), timing observations under proposals DDT5\_003 (PI: Pleunis) and LT5\_003 (PI: Verbiest). Pulsar research at the Jodrell Bank Centre for Astrophysics and the observations using the Lovell Telescope are supported by a consolidated grant from the STFC in the UK. The Nan\c{c}ay Radio Observatory is operated by the Paris Observatory, associated with the French Centre National de la Recherche Scientifique (CNRS). The \textit{Fermi}-LAT Collaboration acknowledges generous ongoing support from a number of agencies and institutes that have supported both the development and the operation of the LAT as well as scientific data analysis. These include the National Aeronautics and Space Administration and the Department of Energy in the United States, the Commissariat \`{a} l'Energie Atomique and the Centre National de la Recherche Scientifique / Institut National de Physique Nucl\'{e}aire et de Physique des Particules in France, the Agenzia Spaziale Italiana and the Istituto Nazionale di Fisica Nucleare in Italy, the Ministry of Education, Culture, Sports, Science and Technology (MEXT), High Energy Accelerator Research Organization (KEK) and Japan Aerospace Exploration Agency (JAXA) in Japan, and the K.~A. Wallenberg Foundation, the Swedish Research Council and the Swedish National Space Board in Sweden. Additional support for science analysis during the operations phase is gratefully acknowledged from the Istituto Nazionale di Astrofisica in Italy and the Centre National d'Etudes Spatiales in France. This work performed in part under DOE Contract DE-AC02-76SF00515.

\vspace{5mm}

\facilities{LOFAR, \textit{Fermi}, GBT, Lovell, NRT}
\software{cdmt, PRESTO, PSRCHIVE, Astropy}

\end{document}